\documentclass[12pt,a4paper]{article}
\pdfoutput=1
\usepackage{jheppub}
\usepackage[utf8]{inputenc}
\usepackage{amsmath,amssymb}
\usepackage{float}
\usepackage{hyperref}

\newcommand\be{\begin{equation}}
\newcommand\ee{\end{equation}}
\newcommand\bea{\begin{eqnarray}}
\newcommand\eea{\end{eqnarray}}

\begin{document}

\begin{titlepage}

\vspace*{1.0cm}

\begin{center}
{\textbf{\huge Quantum Gravity on a Manifold with Boundaries: Schr\"{o}dinger Evolution and Constraints
}}
\end{center}
\vspace{1.0cm}

\centerline{
\textsc{\large J. A.  Rosabal}
\footnote{j.alejandro.rosabal@gmail.com}
}

\vspace{0.6cm}

\begin{center}
{\it Departamento de Electromagnetismo y Electr\'onica,Universidad de Murcia,
Campus de Espinardo, 30100 Murcia, Spain.}\\
{\it Institute for Theoretical and Mathematical Physics, Lomonosov Moscow State University Leninskie Gory, GSP-
1, Moscow, 119991, Russian Federation.}
\end{center}

\vspace*{1cm}

\centerline{\bf Abstract}

In this work, we derive the boundary Schr\"{o}dinger (functional) equation for the wave function of a quantum gravity system on a manifold with boundaries together with a new constraint equation defined on the timelike boundary. From a detailed analysis of the gravity boundary condition on the spatial boundary, we find that while the lapse and the shift functions are independent Lagrange multipliers on the bulk, on the spatial boundary, these two are related; namely, they are not independent. In the Hamiltonian ADM formalism, a new Lagrange multiplier, solving the boundary conditions involving the lapse and the shift functions evaluated on the spatial boundary, is introduced. The classical equation of motion associated with this Lagrange multiplier turns out to be an identity when evaluated on a classical solution of Einstein's equations. On the other hand, its quantum counterpart is a constraint equation involving the gravitational degrees of freedom defined only on the boundary. This constraint has not been taken into account before when studying the quantum gravity Schr\"{o}dinger evolution on manifolds with boundaries.

\begin{centerline}
\noindent

\end{centerline}
\thispagestyle{empty}
\end{titlepage}

\setcounter{footnote}{0}

\tableofcontents

\newpage

\section*{Introduction}\label{intro}

There has recently been a renewed interest in studying manifolds with spatial boundaries in quantum gravity (QG). For instance, in the study of the black hole evaporation \cite{Almheiri:2019hni, Penington:2019kki, Almheiri:2019qdq}, \cite{Giddings:2021ipt}  some models feature a manifold with spatial boundary and with dynamical gravity attached to a nongravitating spacetime. Black hole evaporation is a time dependent process and, for instance, a resolution of the information paradox \cite{Hawking:1976ra} will come with a proper calculation of the quantum amplitude (wave function) for this process to occur. This is why it is imperative to properly understand the time evolution in QG.

 Nowadays, it is well known that on a manifold with spatial boundaries a sort of boundary Schr\"{o}dinger (functional) equation can be derived as a consequence of the presence of this timelike boundary \cite{Hayward:1992ix, Hayward:1993my, Feng:2017xsh}. In this work, we re-derive the QG boundary Schr\"{o}dinger equation on a manifold with boundaries, and we derive a new constraint equation over the spatial boundary  not taken into account before in the literature. Special attention is paid to the boundary conditions on the three-metric induced on the timelike boundary. These boundary conditions link degrees of freedom that on the bulk are independent. The variations over the spatial boundary of these related degrees of freedom are considered in the Lagrangian formulation of gravity. However, in the Hamiltonian formulation,   we show that their variations have not been considered before in the literature. This subtle fact has not been pointed out previously, which implies that we have been doing QG on manifolds with boundaries with a missing equation so far.

Classically, from the Hamiltonian form \cite{Hayward:1992ix,Hawking:1996ww, Brown:2000dz} of the action, we find a new boundary equation of motion. Moreover, we show that this new equation turns out to be an identity when evaluated on a solution of Einstein's equations. The fact that this boundary equation is an identity should be at least gratifying since it is well known that from the variation of the gravity action in the Lagrangian form, no boundary equation appears. So, no contradiction arises between both formulations.

Quantum mechanically, the situation is quite different. In addition to the usual equations (constraints) in the bulk \cite{DeWitt:1967yk}, we find, as the quantum counterpart of the classical boundary equation (identity),  a new boundary constraint, similar to the Wheeler-DeWitt equation in the bulk. This new constraint, together with the boundary Schr\"{o}dinger equation, rules the time evolution of the QG wave function.

The paper is organized as follows.  Section \ref{sec:2} introduces the gravity action on a manifold with spacelike and timelike boundaries. We emphasize the boundary contribution as well as the junction contribution to the action. Finally, we briefly comment on its variation and Einstein's equations.

 In section  \ref{sec:3} we introduce the ADM \cite{Arnowitt:1959eec, Arnowitt:1959ah, Arnowitt:1960es} Lagrangian and Hamiltonian  form of the gravity action, carefully tracking the boundary contributions. We present a detailed analysis of the boundary conditions. The novelty and the key point of this work is based on the observation that in the ADM formulation of gravity on manifolds with boundaries,  when the induced three-metric is fixed on the timelike boundary some  of the Lagrange multipliers that used to be independent over the bulk are linked over the spatial boundary. A new field is introduced to take care of the variation of the linked degrees of freedom over the spatial boundary. We show that its corresponding equation of motion is an identity when evaluated on a solution of Einstein's equations.

 In section \ref{sec:4} we move to the study of quantum gravity and further explore the consequences of the observation previously mentioned. First, we derive the constraint equations on the bulk as well as the Schr\"{o}dinger equation induced by the spatial boundary.  Then we show that due to the fact the boundary conditions do not completely fix the lapse and the shift functions over the spatial boundary, and because these functions are linked over this boundary, a new constraint equation arises.

 Conclusions are presented in section \ref{sec:5} followed by three appendices with some helpful information and details of the calculations carried out along the paper.

\section{Einstein Hilbert Action, Surface and Junction Terms}\label{sec:2}

\begin{figure}[ht]
\centering
\includegraphics[width=.7\textwidth]{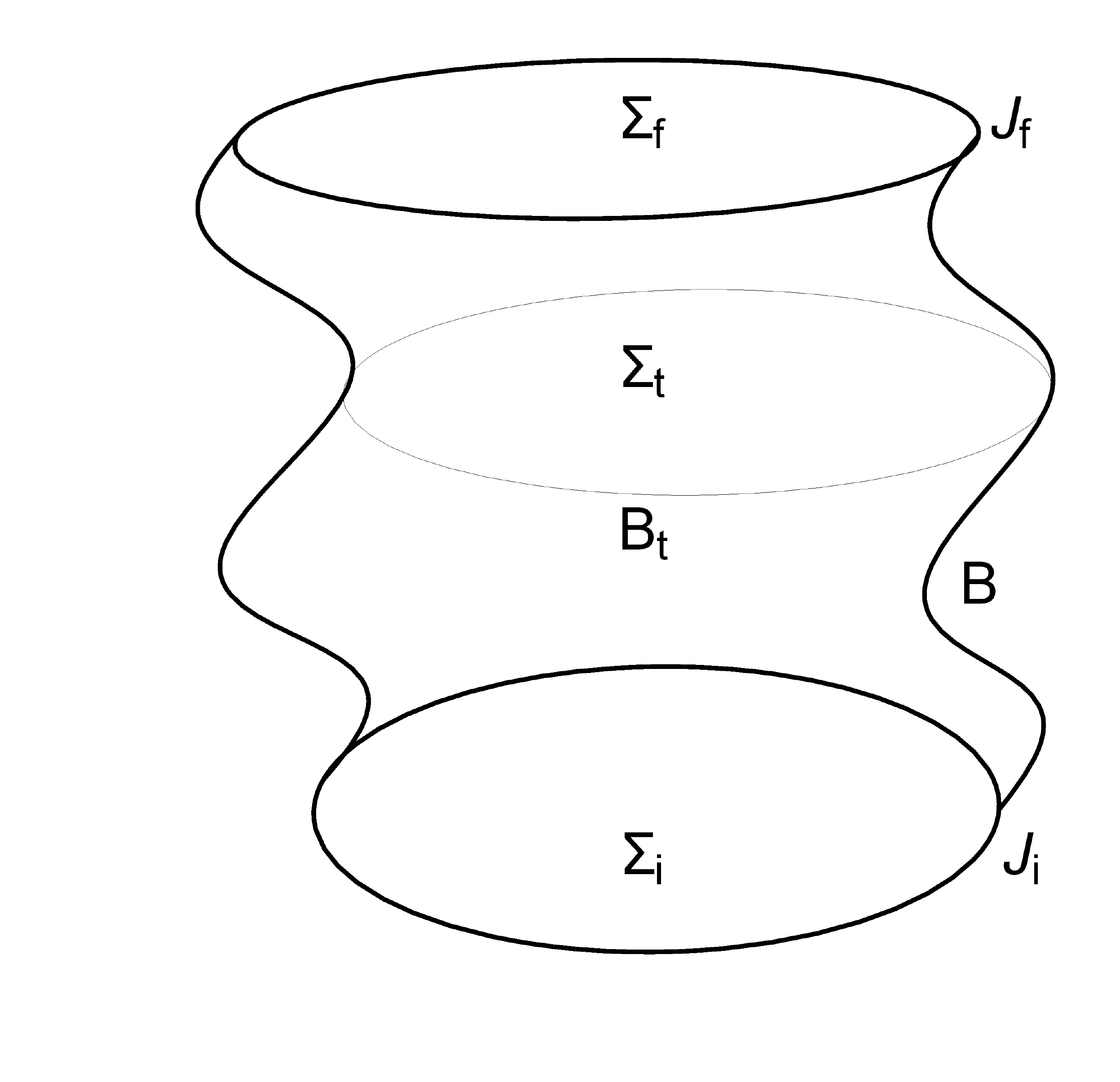}
\caption{\sl The spacetime manifold \text{M}, the spacelike boundaries $\Sigma_{i}$, and $\Sigma_{f}$, the timelike boundary $\text{B}$, and the junctions $\text{J}_i$, and $\text{J}_f$.}\label{fig1}
\end{figure}

Let $\text{M}$, be a sufficiently well behaved four-dimensional spacetime, admitting
a  time function $\text{t}$, which foliates $\text{M}$, into a set of constant $\text{t}$,
spacelike surfaces $\Sigma_{\text{t}}$, \cite{Gourgoulhon:2007ue}. The boundary of $\text{M}$, consists of the
initial and final spacelike surfaces $\Sigma_{i}$, and $\Sigma_{f}$, see Fig. \ref{fig1},  as well as a timelike boundary $\text{B}$. For every $\Sigma_{\text{t}}$, we can define a two surface,
$\text{B}_{\text{t}}=\Sigma_{\text{t}}\cap \text{B}$, which bounds $\Sigma_{\text{t}}$. The set of two surfaces  $\text{B}_{\text{t}}$,  then foliates the boundary $\text{B}$. The intersections  $\Sigma_{i}\cap\text{B}$, and $\Sigma_{f}\cap\text{B}$, are called the junctions and are denoted by $\text{J}_i$, and $\text{J}_f$, respectively.

Nowadays, it is well known that the appropriate form of the Einstein Hilbert action to make quantum gravity on manifolds with non smooth boundaries  Fig. \ref{fig1} includes boundary contributions  \cite{Gibbons:1976ue,York:1986lje} as well as junction contributions \cite{Hayward:1992ix}.
When we fix only the induced three-metric on the boundaries \footnote{Physically this is the only data we are allowed to fix over the boundaries.} the Einstein Hilbert action \cite{Hayward:1992ix,Hawking:1996ww} reads as
\begin{align}\label{totalaction}
  \text{S} = & \frac{1}{16 \pi} \int\limits_{\text{M}}\text{d}^4x \sqrt{-\text{g}}\ \text{R}+ \frac{1}{8 \pi} \int\limits_{\Sigma_f-\Sigma_i}\text{d}^3y \sqrt{\text{h}}\ \text{K} \\
  {} & -\frac{1}{8 \pi} \int\limits_{\text{B}}\text{d}^3\overline{y} \sqrt{-\overline{\text{h}}}\ \overline{\text{K}}+\frac{1}{8 \pi} \int\limits_{\text{J}_f-\text{J}_i}\text{d}^2z \sqrt{\overline{\sigma}}\ \eta
  \nonumber \\
  {} & +\frac{1}{8 \pi} \int\limits_{\text{B}}\text{d}^3\overline{y} \sqrt{-\overline{\text{h}}}\ \overline{\text{K}}_{0} \nonumber.
\end{align}

Where $\text{h}_{ab}$, and $\overline{\text{h}}_{ab}$, are fixed over $\Sigma_i$, $\Sigma_f$, and $\text{B}$. The symbol $\int\limits_{\Sigma_f-\Sigma_i}$, is a short hand notation for $\int\limits_{\Sigma_f}-\int\limits_{\Sigma_i}$. Notice that the unbarred quantities are referred to the spacelike  boundaries $\Sigma_i$, and $\Sigma_f$, while the barred ones to the timelike boundary $\text{B}$. $\text{K}$, and $\overline{\text{K}}$, are the extrinsic curvatures (see appendix \ref{appen:B}) of the boundaries when seen as embedded in the four manifold, and $\text{J}_i$, and $\text{J}_f$, denote the junctions. The metric $\overline{\sigma}_{ij}$, is fixed over $\text{B}$, and in particular over $\text{J}_i$, and $\text{J}_f$,  and
\be
\eta=\text{arcsinh}\big(\hat{\text{n}}^{(\Sigma)}_{|_{\text{B}}}\cdot \hat{\text{n}}^{(\text{B})} \big),\label{eta}
\ee
  where $\hat{\text{n}}^{(\Sigma)}_{|_{\text{B}}}$,  represents the four-dimensional unit normal to  $\Sigma_{t}$, evaluated on $\Sigma_{t}\cap \text{B}$, and  $\hat{\text{n}}^{(\text{B})}$, is the four-dimensional unit normal to $\text{B}$. We would like to emphasize that the scalar field $\eta$ is defined over the entire boundary $\text{B}$,(see for instance appendix \ref{appen:C}) including its boundaries, i.e., the junctions $\text{J}_i$, and $\text{J}_f$, and $\eta\in(-\infty,+\infty)$. The third line in \eqref{totalaction} refers to the quantities computed from the data of the reference background, where the metric is given and we will consider that it is a solution to the Einstein's equations.

We would like to stress that over the boundaries the metric $\text{g}_{\mu\nu}$, is not completely fixed. The only combinations that are fixed are
\be
\text{h}_{ab}=\text{g}_{\mu\nu}\partial_a X_{(\Sigma)}^{\mu}\partial_b X_{(\Sigma)}^{\mu}, \ \ \ \text{and} \ \ \ \overline{\text{h}}_{ab}=\text{g}_{\mu\nu}\partial_a X_{(\text{B})}^{\mu}\partial_b X_{(\text{B})}^{\mu},
\ee
where $X_{(\Sigma)}$, and $X_{(\text{B})}$, are the embedding functions of the boundary surfaces in the four-dimensional spacetime and $a=1,2,3$.

The variation of \eqref{totalaction} is given by \cite{Hayward:1992ix,York:1986lje} (see \cite{Saharian:2003dr} for a detailed presentation about the variations)
\begin{align}\label{variationtotalaction}
  \delta \text{S}& =\frac{1}{16 \pi} \int\limits_{\text{M}}\text{d}^4x \sqrt{-\text{g}}\big( \text{R}_{\mu\nu}-\frac{1}{2}\text{R}\ \text{g}_{\mu\nu}\big)\delta\text{g}^{\mu\nu} \\
  {} & +\frac{1}{16 \pi} \int\limits_{\Sigma_f-\Sigma_i}\text{d}^3y \sqrt{\text{h}}\big( \text{K}^{ab}-\text{K}\ \text{h}^{ab}\big)\delta\text{h}_{ab} \nonumber \\
  {} & -\frac{1}{16 \pi} \int\limits_{\text{B}}\text{d}^3\overline{y} \sqrt{-\overline{\text{h}}}\big( \overline{\text{K}}^{ab}-\overline{\text{K}}\ \overline{\text{h}}^{ab}\big)\delta\overline{\text{h}}_{ab}\nonumber.
\end{align}
Following $\delta\text{S}=0$, and $\delta\text{h}_{ab}=\delta\overline{\text{h}}_{ab}=0$, on the boundaries, we get the vacuum  Einstein's equations of motion
\be\label{eom}
\text{R}_{\mu\nu}-\frac{1}{2}\text{R}\ \text{g}_{\mu\nu}=0.
\ee
Notice that no term including $\eta$, appears in \eqref{variationtotalaction}. In fact the term containing $\eta$, in  \eqref{totalaction} has been included to cancel out a similar term including $\delta \eta$, arising as junctions contributions in the total variation \eqref{variationtotalaction}.

\section{ADM actions, Surface Terms and Boundary Conditions}\label{sec:3}

In this section, we introduce the ADM decomposition of the four-metric. Without loss of the generality, we introduce a foliation over $\Sigma_{\text{t}}$, as well as coordinates such that the spatial boundary is located at a constant value of a coordinate that we call $\text{r}$.  We perform a detailed analysis of the boundary conditions, and we find that another foliation is induced on
$\text{B}$. More importantly, we find the relation that the lapse $\text{N}$, and the shift $\text{N}^a$, Lagrange multipliers satisfy over the spatial boundary. This relation evidences that while $\text{N}$, and $\text{N}^a$, are independent over \text{M}, over \text{B}, they are entangled.

In the Lagrangian formulation, this boundary relation between $\text{N}$, and $\text{N}^a$, is taken into account during the variation process of \eqref{totalaction}. However, we show that in the Hamiltonian formulation, this is not the case. Therefore, to take into account the variations of $\text{N}$, and $\text{N}^a$,  over the spatial boundary it is convenient to introduce a new field to parameterize the boundary conditions.

We present the Lagrangian and the Hamiltonian form of the gravity action in the ADM formalism and the boundary equation of motion associated with the new Lagrange multiplier. We discuss that while the variation of \eqref{totalaction} does not lead to any boundary equation, the variation of the Hamiltonian action does lead to a boundary equation. Nonetheless, we show that this boundary equation of motion is an identity when evaluated on a classical solution of Einstein's equations.

Before discussing the boundary conditions seen from the ADM foliation, we would like to remind the reader that we are assuming that the only data we know on the boundaries is the induced three-metric and this is the only tensor we are allowed to fix over the boundaries.

The ADM metric is given by
\begin{align}\label{adm_metric}
  \text{ds}^2 & =-\text{N}^2\text{dt}^2 +\text{h}_{ab}(\text{dx}^a+\text{N}^a\text{dt})(\text{dx}^b+\text{N}^b\text{dt}) \\
  {} & = -(\text{N}^2-\text{h}_{ab}\text{N}^a\text{N}^b)\text{dt}^2+2\text{h}_{ab}\text{N}^a\text{dx}^b\text{dt}+\text{h}_{ab}\text{dx}^a\text{dx}^b.\nonumber
\end{align}
On a surface of constant time $\Sigma_{\text{t}_{0}}$, i.e., $\text{X}_{(\Sigma)}=(\text{t}_{0},\text{x}^1,\text{x}^2,\text{x}^3)$, the metric takes the form
\be
\text{ds}^2_{|_{\text{t}_{0}}}=\text{h}_{ab}\text{dx}^a\text{dx}^b.
\ee
At this point we find convenient to introduce another foliation. Introducing a set of coordinates on $\Sigma_{\text{t}}$, $(\text{x}^1,\text{x}^2,\text{x}^3)=(\text{r},\text{x}^2,\text{x}^3)$,  such that the location of $\text{B}$, is at a constant value of the coordinate,  $\text{r}=\text{r}_{\text{cons}}$. We can write the three-dimensional metric $\text{h}_{ab}$,  on $\Sigma_{\text{t}_{0}}$, as
\be
\text{ds}^2_{|_{\text{t}_{0}}}=\Lambda^2\text{dr}^2+\sigma_{ij}(\text{dx}^i+\Lambda^i \text{dr})(\text{dx}^j+\Lambda^j \text{dr}),
\ee
where $i=2,3$.
In this coordinates we just have made the identification $\text{h}_{11}=\text{h}_{\text{rr}}=\Lambda^2+\sigma_{ij}\Lambda^i\Lambda^j$, and $\text{h}_{1i}=\text{h}_{\text{r}i}=\sigma_{ij}\Lambda^j$. Now, at a constant time surface and  at $\text{r}=\text{r}_{\text{cons}}$, the metric takes the form
\be
\text{ds}^2_{|_{(\text{t}_{0},\text{r}_{\text{cons}})}}=
\sigma_{ij}{}_{|_{\text{B}_{\text{t}_{0}}}}\text{dx}^i\text{dx}^j,
\ee
with the boundary  metric  $\sigma_{ij}{}_{|_{\text{B}_{\text{t}_{0}}}}=\overline{\sigma}_{ij}$,  fixed;   $\text{B}_{\text{t}_{0}}$, denotes $\Sigma_{\text{t}_{0}}\cap \text{B}$.

Now, the induced three-metric on $\text{B}$, takes the form
\begin{align}\label{inducemetrB}
  \text{ds}^2_{|_{\text{r}_{\text{cons}}}} & =-\Big(\text{N}^2 -(\Lambda \text{N}^{\text{r}} )^2-\overline{\sigma}_{ij}(\Lambda^i \text{N}^{\text{r}} +\text{N}^i)(\Lambda^j \text{N}^{\text{r}}+\text{N}^j)\Big)\text{dt}^2 \\
  {} & +2\overline{\sigma}_{ij}(\Lambda^i \text{N}^{\text{r}} +\text{N}^i)\text{dx}^j\text{dt}+\overline{\sigma}_{ij}\text{dx}^i\text{dx}^j.
\end{align}
Which can be rewritten as
\be\label{inducemetrB fol}
  \text{ds}^2_{|_{\text{r}_{\text{cons}}}}  =-\Big(\text{N}^2 -(\Lambda \text{N}^{\text{r}} )^2\Big)\text{dt}^2
  +\overline{\sigma}_{ij}\Big(\text{dx}^i+(\Lambda^i \text{N}^{\text{r}} +\text{N}^i)\text{dt}\Big)\big(\text{dx}^j+(\Lambda^j \text{N}^{\text{r}} +\text{N}^j)\text{dt}\Big),
\ee
making explicitly a third foliation, this time on $\text{B}$.
Notice that in addition to $\overline{\sigma}_{ij}$, the combinations
\be\label{parametrization1}
\text{N}^2 -(\Lambda \text{N}^{\text{r}} )^2=\overline{\text{N}}^2,
\ee
and
\be\label{parametrization2}
(\Lambda^i \text{N}^{\text{r}} +\text{N}^i)=\overline{\text{N}}^i,
\ee
must be fixed over $\text{B}$. So,  $\overline{\text{N}}$, $\overline{\text{N}}^i$, and $\overline{\sigma}_{ij}$,  are given functions of $(t,\text{x}^2,\text{x}^3)$, over $\text{B}$,  i.e., the three-metric
\be\label{inducemetrB foln}
  \text{ds}^2_{|_{\text{r}_{\text{cons}}}}  =-\overline{\text{N}}^2\text{dt}^2
  +\overline{\sigma}_{ij}(\text{dx}^i+\overline{\text{N}}^i\text{dt})\big(\text{dx}^j+\overline{\text{N}}^j\text{dt}),
\ee
should be completely specified on $\text{B}$.

\subsection{Unit normal vectors}\label{sec:3.1}

In this section we shall  present the unit normal vectors to the surfaces $\Sigma_t$, and $\text{B}$. Their expressions will be helpful for the subsequent discussions.  In the coordinates $(t,\text{r},\text{x}^1,\text{x}^2)$, one can define the functions
\bea
\text{F}^{(\Sigma)} & = & t_{0}-t, \nonumber\\
\text{F}^{(\text{B})} & = & \text{r}_{0}-\text{r}.
\eea
Such that a surface of constant time $t_0$, or constant $\text{r}$,  $\text{r}_0$, (in particular the boundary surfaces) are simply specified by the relations $\text{F}^{(\Sigma)}=\text{F}^{(\text{B})}= 0$.

The unit normal vectors are defined as
\bea
\hat{\text{n}}_{\mu}^{(\Sigma)} & = & \frac{\partial_{\mu}\text{F}^{(\Sigma)}}{(-\text{g}^{\alpha\beta}\partial_{\alpha}\text{F}^{(\Sigma)}\partial_{\beta}\text{F}^{(\Sigma)})^{\frac{1}{2}}}=\text{N}(-1,0,0,0),\\
\hat{\text{r}}_{\mu} & = & \frac{\partial_{\mu}\text{F}^{(\text{B})}}{(\text{g}^{\alpha\beta}\partial_{\alpha}\text{F}^{(\text{B})}\partial_{\beta}\text{F}^{(\text{B})})^{\frac{1}{2}}}=
\frac{1}{(\Lambda^{-2}-\text{N}^{-2}\text{N}^{\text{r}}\text{N}^{\text{r}})^{\frac{1}{2}}}(0,-1,0,0),
\eea
with $\hat{\text{r}}_{|_{\text{r}_{\text{cons}}}}=\hat{\text{n}}^{(\text{B})}$.

The scalar product between both unit normal vectors over $\text{B}$ is
\be
(\text{g}^{\mu\nu}\hat{\text{n}}_{\mu}^{(\Sigma)}\hat{\text{n}}_{\nu}^{(\text{B})})_{|_{\text{r}_{\text{cons}}}}=\frac{\Lambda\text{N}^{\text{r}}}{(\text{N}^2-(\Lambda \text{N}^{\text{r}})^2)^{\frac{1}{2}}}
=\frac{\Lambda\text{N}^{\text{r}}}{\overline{\text{N}}
}\label{scalar_prod},
\ee
where the matrix form of the metrics involved  in this calculation can be found in appendix \ref{appen:A}. Notice that now \eqref{eta} can be written as
\be
\eta=\text{arcsinh}\big(\frac{\Lambda\text{N}^{\text{r}}}{\overline{\text{N}}} \big).\label{eta1}
\ee

\subsection{Boundary conditions}\label{sec:3.2}

Before continuing we shall discuss the boundary conditions expressed in the coordinates $(\text{t}, \text{r},\text{x}^2,\text{x}^3)$, for the system we are interested in.

We would like to stress that while $\sigma_{ij}$, is fixed over $\text{B}$, the rest of the degrees of freedom involved in relations \eqref{parametrization1} and \eqref{parametrization2} are not fixed over $\text{B}$. Namely, we can not impose separately conditions on each functions appearing in these relations. For instance, according to the boundary conditions \eqref{parametrization1} and \eqref{parametrization2} we can not separately set $\text{N}^{\text{r}}=0$, and $\text{N}=\overline{\text{N}}$, over $\text{B}$. Actually, \eqref{parametrization1}  has much more solutions than $\text{N}^{\text{r}}=0$, and $\text{N}=\overline{\text{N}}$. In fact it has infinitely many solutions.

Using \eqref{eta1} we note that
\bea\label{eta_def1}
\Lambda \text{N}^{\text{r}}_{|_{\text{r}_{\text{cons}}}} & = & \overline{\text{N}}\ \text{sinh}(\eta),\\
\text{N}_{|_{\text{r}_{\text{cons}}}} & = & \overline{\text{N}}\ \text{cosh}(\eta) \nonumber.
\eea
The previous expressions indicate that the infinitely many solutions can be parameterized \footnote{Although \eqref{parametrization1} can be parameterized in infinitely many ways, we find convenient to use \eqref{eta_def1} because it  is the optimum one to carry out the subsequent calculations.} using the $\eta$ field introduced in section \ref{sec:2}.

\subsection{ADM action, Hamiltonian and Surface Terms}\label{sec:2.3}
In this section we derive the ADM form of the gravity action, paying special attention to the boundary contribution. We refer the reader to appendix \ref{appen:C} for some details of the calculation presented here, see also \cite{Hawking:1996ww}.  Using
\be\label{R3}
\text{R}={}^{(3)}\text{R}+\text{K}_{ab}\text{K}^{ab}-\text{K}^2+2\nabla_{\mu}(-\text{K}\hat{\text{n}}^{\mu}-\hat{\text{n}}^{\nu}\nabla_{\nu}\hat{\text{n}}^{\mu}),
\ee
and
\be
{}^{(2)}\text{K}=-\text{g}^{\mu\nu}\nabla_{\mu}{}^{(3)}\hat{\text{r}}_{\nu}+\hat{\text{n}}^{\mu}\nabla_{\mu}\hat{\text{n}}_{\nu}{}^{(3)}\hat{\text{r}}^{\nu}.
\ee
Where ${}^{(2)}\text{K}$, is the extrinsic curvature of the two-dimensional surface $\text{B}_{\text{t}}$, as embedded in $\Sigma_{\text{t}}$,  and ${}^{(3)}\hat{\text{r}}_{\nu}=\Lambda(-\text{N}^\text{r},-1,0,0)$,  its unit three-dimensional normal vector embedded in four dimensions, see appendix \ref{appen:B}, and $\hat{\text{n}}_{\mu}=\text{N}(-1,0,0,0)$, the unit normal to $\Sigma_{\text{t}}$.

After plugging \eqref{R3} in \eqref{totalaction} the total derivatives can be integrated to give rise new boundary terms. This process introduces the four-dimensional unit normal vector to $\text{B}$, $\hat{\text{r}}_{\mu}$, given by
\be\label{refcapri1}
\hat{\text{r}}_{\mu}=\text{M}^{-1}\Lambda\text{N}(0,-1,0,0),
\ee
where $\text{M}=\big(\text{N}^2-(\Lambda\text{N}^{\text{r}})^2\big)^{\frac{1}{2}}$. We will explicitly keep the term $\text{M}$, which on \text{B}, $\text{M}_{|_{\text{r}_{\text{cons}}}}=\overline{\text{N}}$, because in the subsequent discussion it will appear under derivative with  respect to $\text{r}$, in which case we need to take into account  its definition away from $\text{B}$.
Using also the relation among the normal vectors
\be\label{refcapri2}
\hat{\text{r}}_{\mu}=\text{M}^{-1}\big(\text{N}{}^{(3)}\hat{\text{r}}_{\mu}-\Lambda\text{N}^{\text{r}}\hat{\text{n}}_{\mu} \big),
\ee
and the boundary conditions \eqref{parametrization1} and \eqref{parametrization2}
the action \eqref{totalaction} can be rewritten in the ADM form
\begin{align}\label{ADMaction}
  \text{S} & = \int\limits_{\text{M}}\text{d}^4x \ {\cal L}\nonumber\\
  {} & = \frac{1}{16 \pi} \int \text{dt} \int\limits_{\Sigma_{\text{t}}}\text{d}^3x \text{N}\sqrt{\text{h}}\Big( \text{K}_{ab}\text{K}^{ab}-\text{K}^2+ {}^{(3)}\text{R}\Big)\nonumber \\
  {} & - \frac{1}{8 \pi} \int\text{dt}\int\limits_{\text{B}_{\text{t}}}\text{dx}^2 \text{N} \sqrt{\overline{\sigma}}\ {}^{(2)}\text{K}
+\frac{1}{8 \pi} \int\text{dt}\int\limits_{\text{B}_{\text{t}}}\text{dx}^2 \big(\partial_{\text{t}}(\sqrt{\overline{\sigma}})-\partial_j(\sqrt{\overline{\sigma}} \ \overline{\text{N}}^j) \big)\ \eta.
\end{align}
Where  $\text{K}_{ab}$, and ${}^{(2)}\text{K}$,  are given by
\be
\text{K}_{ab}=-\text{N}^{-1}\Big(\frac{1}{2}\partial_{t}\text{h}_{ab}-\text{D}_{(a}\text{N}_{b)} \Big)\quad;\quad \text{K}=\text{h}^{ab}\text{K}_{ab}\label{K},
\ee
and
\be
{}^{(2)}\text{K}_{ij}=\Lambda^{-1}\Big(\frac{1}{2}\partial_{\text{r}}\sigma_{ij}-{}^{(2)}\text{D}_{(i}\Lambda_{j)} \Big)\quad;\quad {}^{(2)}\text{K}=\overline{\sigma}^{ij}{}^{(2)}\text{K}_{ij} \label{2K},
\ee
with $\text{D}_a$, and ${}^{(2)}\text{D}_i$,  the covariant derivatives compatible with $\text{h}_{ab}$, and  $\overline{\sigma}_{ij}$, respectively; and
\bea\label{determinats}
\sqrt{\text{h}} & = & \text{N}\sqrt{\sigma},\\
\sqrt{-\overline{\text{h}}} & = & \overline{\text{N}}\sqrt{\overline{\sigma}}\nonumber.
\eea
Notice that in the last term in \eqref{ADMaction}
\be\label{extrincBtinB}
\big(\partial_{\text{t}}(\sqrt{\overline{\sigma}})-\partial_j(\sqrt{\overline{\sigma}} \ \overline{\text{N}}^j) \big)=
\sqrt{-\overline{\text{h}}} \big(\overline{\text{D}}_{\alpha}\text{v}^{\alpha} \big)= -\sqrt{-\overline{\text{h}}}\ {}^{(2)}\overline{\text{K}},
\ee
where $\overline{\text{D}}_{\alpha}$, is the covariant derivative compatible with the metric \eqref{inducemetrB foln} on $\text{B}$, and $\text{v}^{\alpha}$, the unit normal vector to $\text{B}_{\text{t}}$, given by
\be
\text{v}^{\alpha}=\overline{\text{N}}^{\ -1}\Big(1,-\overline{\text{N}}^j\Big),
\ee
and ${}^{(2)}\overline{\text{K}}$, the extrinsic curvature of $\text{B}_{\text{t}}$, as embedded in $\text{B}$.

Defining the canonically conjugate momenta
\be \label{momenta def0}
\Pi^{ab}=\frac{\partial{\cal L}}{\partial\dot{\text{h}}_{ab}}=(16\pi)^{-1}\sqrt{\text{h}}(-\text{K}^{ab}+\text{h}^{ab}\text{K}),
\ee
the Hamiltonian form of the ADM action \eqref{ADMaction} reads as
\begin{align}\label{ADMactionH}
  \text{S} & = \text{S}_{\text{M}}+\text{S}_{\text{B}}\nonumber\\
  \text{S} & =  \int \text{dt} \int\limits_{\Sigma_{\text{t}}}\text{d}^3x \Big( \Pi^{ab}\partial_{\text{t}} \text{h}_{ab}-\text{N}{\cal H}-\text{N}_a{\cal H}^a \Big)\nonumber \\
  {} & -  \int\text{dt}\int\limits_{\text{B}_{\text{t}}}\text{dx}^2\sqrt{\overline{\sigma}}\Big(2{}^{(3)}\hat{\text{r}}_a\frac{\Pi^{ab}}{\sqrt{\text{h}}}\text{N}_b+(8\pi)^{-1}\text{N} {}^{(2)}\text{K}+(8\pi)^{-1} \overline{\text{N}}{}^{(2)}\overline{\text{K}}\  \eta\Big).
\end{align}
Where
\be
{\cal H}=\frac{16\pi}{\sqrt{\text{h}}}(\Pi_{ab}\Pi^{ab}-\frac{1}{2}\Pi^2)-\frac{\sqrt{\text{h}}}{16\pi}{}^{(3)}\text{R},
\ee
and
\be
{\cal H}^a=-2\text{D}_b\Pi^{ba},
\ee
  with ${}^{(3)}\hat{\text{r}}_a$, the unit normal (with respect to the three-metric $\text{h}_{ab}$) to \text{B}. Using the chain rule and \eqref{momenta def0} we can express  the momenta $\Pi^{ab}$, in the coordinates $(\text{t},\text{r},\text{x}^2,\text{x}^3)$,
  \bea \label{momenta def}
  \Pi^{11} & = & \frac{1}{2}\Lambda^{-1} \text{P}, \\
  \Pi^{1i} & = &\frac{1}{2}(\text{P}^i-\Lambda^{-1}\Lambda^{i}\text{P}), \nonumber\\
  \Pi^{ij} & = & \text{P}^{ij}+\frac{1}{2}\Lambda^{-1}\Lambda^i\Lambda^j\text{P}-\frac{1}{2}(\text{P}^i\Lambda^j+\text{P}^j\Lambda^i).\nonumber
  \eea
  Where $(\text{P}, \text{P}_i, \text{P}^{ij})$, are the canonically conjugate momenta associated to $(\Lambda,\Lambda^i, \sigma_{ij})$, respectively .
  While ${}^{(3)}\hat{\text{r}}_a$, and $\text{N}_{a|_{\text{r}_{\text{cons}}}}$, in the coordinates $(\text{t},\text{r},\text{x}^2,\text{x}^3)$, read as
\be
{}^{(3)}\hat{\text{r}}_a=\Lambda(-1,0,0),
\ee
and
\be\label{na-on-B}
\text{N}_{a|_{\text{r}_{\text{cons}}}}=(\Lambda^2\text{N}^{\text{r}}+\Lambda_j\overline{\text{N}}^j,\overline{\text{N}}_i),
\ee
and
\be
\Pi^{ab}\partial_{\text{t}} \text{h}_{ab}=\text{P}\partial_{\text{t}}\Lambda+\text{P}_i\partial_{\text{t}}\Lambda^i+\text{P}^{ij}\partial_{\text{t}}\sigma_{ij}.
\ee

In the Hamiltonian formulation of gravity $\text{N}$, and $\text{N}^a$, play the role of Lagrange multipliers \cite{Hartle:1983ai}. Their equations of motion only enforce the Hamiltonian and the momentum constraint ${\cal H}={\cal H}^a=0$, on $\text{M}$.

Variations of \eqref{ADMactionH} with respect to $\text{N}$, and $\text{N}_a$,  and $\text{h}_{ab}$; and $\Pi^{ab}$,  over $\text{M}$,  lead (after some manipulations) to the same equations of motion  \eqref{eom}. However, for the action written in the ADM Hamiltonian form extra care is needed with the variations over $\text{B}$. Notice that while the variations of the Lagrange multipliers $\text{N}$, and $\text{N}^{\text{r}}$, are independent over $\text{M}$, over the boundary $\text{B}$,  these functions are entangled each other through the relations \eqref{parametrization1} and \eqref{parametrization2}. So, the variation of $\text{N}$, and $\text{N}^{\text{r}}$, over $\text{B}$, are not independent.

To take care of the variations  of  $\text{N}$, and $\text{N}^{\text{r}}$, over $\text{B}$, it is convenient to parameterize the solutions of \eqref{parametrization1}, using \eqref{eta_def1}, and regard $\eta$, as an independent Lagrange multiplier defined over $\text{B}$. With this choice the boundary action $\text{S}_{\text{B}}$, in the coordinates $(\text{t},\text{r},\text{x}^2,\text{x}^3)$,  can be rewritten as
\begin{align}\label{Baction}
  \text{S}_{\text{B}} =& -  \int\text{dt}\int\limits_{\text{B}_{\text{t}}}\text{dx}^2\Big(- \text{P}\ \overline{\text{N}}\ \text{sinh}(\eta)+(8\pi)^{-1} \sqrt{\overline{\sigma}} {}^{(2)}\text{K}\ \overline{\text{N}}\ \text{cosh}(\eta) \\
  {} & +(8\pi)^{-1}\sqrt{\overline{\sigma}}{}^{(2)}\overline{\text{K}}\ \overline{\text{N}}\ \eta-\text{P}_j\overline{\text{N}}^j\Big)\nonumber.
\end{align}

The boundary equation of motion, associated to $\eta$, on $\text{B}$, i.e., $\frac{\delta }{\delta \eta}\text{S}_{\text{B}}=0$,  takes the form
\be\label{boundary_eom_compact}
\text{P}\ \text{cosh}(\eta)-(8\pi)^{-1} \sqrt{\overline{\sigma}} {}^{(2)}\text{K}\  \text{sinh}(\eta)
 -(8\pi)^{-1}\sqrt{\overline{\sigma}}{}^{(2)}\overline{\text{K}}=0,
\ee
or, using the first relation in \eqref{momenta def}, \eqref{extrincBtinB}, \eqref{determinats} and \eqref{eta_def1},
\be\label{boundary_eom1}
-2\Big(\frac{\Pi^{11}}{\sqrt{\text{h}}}\Big)\Lambda^2\text{N}+(8\pi)^{-1}{}^{(2)}\text{K}\Lambda\text{N}^{\text{r}}-(8\pi)^{-1} \frac{\partial_{\text{t}}\sqrt{\overline{\sigma}}}{\sqrt{\overline{\sigma}}}+(8\pi)^{-1}{}^{(2)}\text{D}_i\overline{\text{N}}^i=0,
\ee
with each function appearing in \eqref{boundary_eom1} evaluated on $\text{B}$.

At this point, it might be disturbing the fact that from general considerations the variation of \eqref{totalaction} does not lead to any boundary equation of motion. However, the variation of \eqref{ADMactionH} does lead to \eqref{boundary_eom1}. It turns out that classically this boundary equation of motion does not play any relevant role, and it is not needed to solve it. Rather than being a new equation, classically, \eqref{boundary_eom1} is an identity. It means that after solving Einstein's equations on $\text{M}$, the boundary equation gets automatically satisfied.

In what follows we show that \eqref{boundary_eom1} is actually an identity.
In the Hamiltonian formulation $\text{h}_{ab}$, and $\Pi^{ab}$, are two set of independent degrees of freedom, however on a solution of the Einstein's equations the momenta verify the equation
\be\label{momenta-pi-ab}
\Pi^{ab}=(16\pi)^{-1}\sqrt{\text{h}}\Big(-\text{K}^{ab}+\text{h}^{ab}\text{K} \Big).
\ee
These equations are obtained upon varying \eqref{ADMactionH} with respect to $\Pi^{ab}$.

Using \eqref{K} (see also appendix \ref{appen:A}), after some algebra we get, that on $\text{B}$,
\be
(16\pi)\Big(\frac{\Pi^{11}}{\sqrt{\text{h}}}\Big)=-\text{K}^{11}+\text{h}^{11}\text{K}=\text{N}^{-1}\Lambda^{-2}\Big(-\frac{1}{2}\overline{\sigma}^{ij}\partial_{\text{t}}\overline{\sigma}_{ij}+\overline{\sigma}^{ij}\text{D}_i\text{N}_j\Big).
\ee
Now, using \eqref{na-on-B} and \eqref{2K} we can show that on $\text{B}$,
\be\label{relation-on-B}
\overline{\sigma}^{ij}\text{D}_i\text{N}_j=\overline{\sigma}^{ij}\Big(\partial_i\overline{\text{N}}_j-\Gamma^k{}_{ij}\overline{\text{N}}_k-\Gamma^1{}_{ij}\text{N}_1 \Big)={}^{(2)}\text{D}_i\overline{\text{N}}^i+{}^{(2)}\text{K}\Lambda\text{N}^{\text{r}},
\ee
Collecting the results \eqref{momenta-pi-ab}-\eqref{relation-on-B} and using
\be\label{derivativeofdet}
\frac{\partial_{\text{t}}\overline{\sigma}}{\overline{\sigma}}=\overline{\sigma}^{ij}\partial_{\text{t}}\overline{\sigma}_{ij},
\ee
one can show that indeed on a classical solution  \eqref{boundary_eom1} is an identity. So, \eqref{boundary_eom1}, which is the equation of motion corresponding to $\eta$, defined only on the spatial boundary, is an identity because it is already contained in the equations \eqref{momenta-pi-ab}. In other words, equation \eqref{boundary_eom1} can be viewed as the restriction to the boundary of the $a=b=1$, component of the equations \eqref{momenta-pi-ab}. We would like to emphasize  that  \eqref{boundary_eom1} and \eqref{momenta-pi-ab} come from the variation of two different and unrelated  degrees of freedom.

 To the knowledge of the author,  this is a new result which has not been presented before in the literature. In the next sections we will see that the quantum counterpart of the boundary equation \eqref{boundary_eom_compact} becomes a constraint equation on $\text{B}$, similar to the Wheeler-Dewitt equation and the momentum constraint on $\text{M}$.

\section{Quantum Gravity on a Manifold with boundaries}\label{sec:4}

In this section, we shall derive the Schr\"{o}dinger equation as well as the constraint equations on $\text{M}$, and a new constraint equation induced by the presence of $\text{B}$. We work formally using the path integral for gravity.

Before continuing we would like to briefly discuss the measure of the gravity path integral (see \cite{Hawking:1996ww} and  \cite{Teitelboim:1981ua}, and references therein) as well as which degrees of freedom are integrated over each of the four regions, namely, $\text{M}$,  $\Sigma_{(i,f)}$, $\text{B}$, and $\text{B}_{(i,f)}=\text{J}_{(i,f)}$. They have been summarized in Table \ref{table666}.

\begin{table}[ht]
\begin{tabular}{ |c|c|c|c|c| }
\hline
                      &$\text{M}$                               &$\Sigma_{(i,f)}$                                         &$\text{B}$              &$\text{B}_{(i,f)}=\text{J}_{(i,f)}$\\ \hline

 Integrated d.o.f     &$(\underbrace{\text{N},\text{N}^a}_{\text{4}},\underbrace{\Lambda,\Lambda^i,\sigma_{ij})}_{\text{6}}$&$\underbrace{(\text{N},\text{N}^a)}_{\text{4}}$&$\underbrace{(\eta,\Lambda,\Lambda^i)}_{\text{4}}$&$\underbrace{\eta}_{\text{1}}$\\ \hline
 Fixed d.o.f          &$\text{none}$&$\underbrace{(\Lambda,\Lambda^i,\sigma_{ij})}_{\text{6}}$&$\underbrace{(\overline{\text{N}},\overline{\text{N}}^i,\overline{\sigma}_{ij})}_{\text{6}}$&  $\underbrace{(\overline{\text{N}},\overline{\text{N}}^i,\Lambda,\Lambda^i,\overline{\sigma}_{ij})}_{\text{9}}$  \\ \hline
\end{tabular}
\caption{\sl Integrated {\it vs} fixed degrees of freedom over each region. Notice that the number of integrated and fixed degrees of freedom add up to ten over each region.}\label{table666}
\end{table}

When writing down the path integral one has to integrate over the degrees of freedom  which are varied in the action  to get the classical equations of motion. To get the Einstein's equation  on $\text{M}$, one has to vary the action with respect to $\text{h}_{ab}\rightarrow(\Lambda,\Lambda^i,\sigma_{ij})$, and $(\text{N},\text{N}^a)$. As $(\Lambda,\Lambda^i,\sigma_{ij})$, are fixed over the spacelike boundaries, the variations of these degrees of freedom  do not extend until $\Sigma_{i}$ and $\Sigma_{f}$.

On the other hand, $(\text{N},\text{N}^a)$, are not fixed on the spacelike boundaries so, their variations extend until $\Sigma_{i}$ and $\Sigma_{f}$. On the timelike boundary $\text{B}$, we fix different combinations of the metric. While on $\Sigma_{i}$ and $\Sigma_{f}$, we fix the three-metric $\text{h}_{ab}\rightarrow(\Lambda,\Lambda^i,\sigma_{ij})$,  on $\text{B}$, we fix the three-metric $\overline{\text{h}}_{ab}\rightarrow(\overline{\text{N}},\overline{\text{N}}^i,\overline{\sigma}_{ij})$. Notice that on $\text{B}$, also six functions are specified. It means that out of the ten degrees of freedom in the metric tensor four remain unfixed on $\text{B}$. The most obvious of the unfixed ones are $(\Lambda,\Lambda^i)$, they add up to three \footnote{ Since $\Lambda^i$, and $\text{N}^i$, are related through the boundary condition \eqref{parametrization2} we could also use as integration variables the functions $(\Lambda, \text{N}^i)$.}. Still one more is missing. As discussed in section \ref{sec:3} to take care of the variations of $(\text{N},\text{\text{N}}^{\text{r}})$, over $\text{B}$, which are not independent but they are related as indicated in \eqref{parametrization1} and \eqref{parametrization2}, we have to introduce a new degree of freedom that we call $\eta$. Its variation leads to a boundary equation that on a classical solution of the Einstein's equations can be shown it is an identity. So, the integrated degrees of freedom over $\text{B}$, are $(\eta,\Lambda,\Lambda^i)$, which now, add up to four.

Only the junctions $\text{J}_{(i,f)}=\Sigma_{(i,f)}\cap \text{B}$, remain to be analyzed. On the one hand, as $\text{J}_{(i,f)}\in\Sigma_{(i,f)}$, on $\text{J}_{(i,f)}$, $(\Lambda,\Lambda^i,\sigma_{ij})$, are fixed. On the other hand as $\text{J}_{(i,f)}\in\text{B}$, in addition to $(\overline{\text{N}},\overline{\text{N}}^i)$, now $(\Lambda,\Lambda^i)$, are fixed too. The only degree of freedom that remains unfixed over the junctions is $\eta$. This unfixed degree of freedom, defined over $\text{B}$, including the junctions, is precisely what allows us to derive a new constraint that has not been considered in the literature before.

Taking into account what has  been exposed in the previous paragraphs we can define the measure of the path integral as $\text{D}\mu=\text{D}\mu_1\text{D}\mu_2$,
where
\be
\text{D}\mu_1=\underset{\text{t}\in[\text{t}_i,\text{t}_f]}{\Pi}\Big[\text{D}\text{P} \ \text{D}\text{P}_i \ \text{D}\text{P}^{ij} \ \text{D}\text{N} \ \text{D}\text{N}^a \Big]_{\big{|}_{\Sigma_{\text{t}}}} \ \underset{\text{t}\in(\text{t}_i,\text{t}_f)}{\Pi}\Big[\text{D}\Lambda \ \text{D}\Lambda^i \ \text{D}\sigma_{ij}\Big]_{\big{|}_{\Sigma_{\text{t}}}},
\ee
and
\be
\text{D}\mu_2=\underset{\text{t}\in[\text{t}_i,\text{t}_f]}{\Pi}\Big[\text{D}\text{P} \ \text{D}\text{P}_i \ \text{D}\text{P}^{ij} \ \text{D}\eta \Big]_{\big{|}_{\text{B}_{\text{t}}}} \ \underset{\text{t}\in(\text{t}_i,\text{t}_f)}{\Pi}\Big[\text{D}\Lambda \ \text{D}\Lambda^i\Big]_{\big{|}_{\text{B}_{\text{t}}}},
\ee
where the brackets in the previous expressions $\Big[\text{...}\Big]$ are understood as the proper measure of the path integral (including gauge fixing terms) together with the product over every point on $\Sigma_{\text{t}}$, or $\text{B}_{\text{t}}$. As we are interested in the Hamiltonian form of the action, we have included the integration in the momenta $(\text{P},\text{P}_i,\text{P}^{ij})$, in the measure.

\subsection{Schrödinger Evolution}

In the path integral formulation we can write the wave function (strictly speaking it is a functional) \cite{Hartle:1983ai} as
\be\label{wf}
\Psi=\int\text{D} \mu\text{e}^{\text{i}\text{S}}.
\ee
On a manifold with spatial boundaries, in addition to the usual arguments of the wave function, i.e., the three-metric $\text{h}_{ab}$, over $\Sigma_{f}$, the functional \eqref{wf} depends also on three-metric $\overline{\text{h}}_{ab}$, the proper time $\overline{\tau}$, and proper space $\overline{\chi}^i$, on the timelike boundary, see \eqref{propers}, that is
\be
\Psi=\Psi\big[\Lambda,\Lambda^i,\sigma_{ij},\overline{\text{N}},\overline{\text{N}}^i,\overline{\sigma}_{ij};\overline{\tau},\overline{\chi}^i\big).
\ee
In this section we assume that the measure is invariant under translation of $\text{N}$, and $\text{N}^a$, over \text{M}; and invariant under translation of $\eta$, over \text{B}. If it is not, there would be, in
addition, a divergent contribution to the relations we will obtain here. This contribution must be suitably regulated to zero or cancel possible divergences arising from the calculation. We also ignore the problem of the operator ordering, although the discussion presented here is compatible with a possible resolution of it  \cite{Hawking:1985bk}.

For the subsequent discussion we find convenient to write the ADM  Hamiltonian action in the form
\begin{align}
\text{S} & =  \int \text{dt} \int\limits_{\Sigma_{\text{t}}}\text{dr}\text{d}^2x \Big(\text{P}\partial_{\text{t}}\Lambda+\text{P}_i\partial_{\text{t}}\Lambda^i+\text{P}^{ij}\partial_{\text{t}}\sigma_{ij}-\text{N}{\cal H}-\text{N}^a{\cal H}_a \Big)\nonumber \\
  {} & + \int\text{dt}\int\limits_{\text{B}_{\text{t}}}\text{dx}^2\Big((8\pi)^{-1}\eta\ \partial_{\text{t}} \sqrt{\overline{\sigma}}-\overline{\text{N}}\ \overline{{\cal H}}-\overline{\text{N}}^j\ \overline{{\cal H}}_j \Big),
\end{align}
where
\bea
\overline{{\cal H}} & = & -\text{P}\ \text{sinh}(\eta)+(8\pi)^{-1}\sqrt{\overline{\sigma}}{}^{(2)}\text{K}\ \text{cosh}(\eta),\nonumber\\
\overline{{\cal H}}_j & = &-\big( \text{P}_j+(8\pi)^{-1}\partial_j\eta\big).
\eea
The variation of $\Psi$, induced by the change of the boundary data $(\Lambda,\Lambda^i,\sigma_{ij})$, on $\Sigma_{f}$, while we push forward this surface, is given by
\begin{align}\label{psi-variation}
\delta\Psi & = \text{i} \int \text{D}\mu \Bigg[ \int\limits_{\Sigma_{f}}\text{dr}\text{d}^2x \Big(\text{P}\delta\Lambda+\text{P}_i\delta\Lambda^i+\text{P}^{ij}\delta\sigma_{ij}\Big)\nonumber \\
{} & -\int\limits_{\Sigma_{f}}\text{dr}\text{d}^2x\Big(\text{N}{\cal H}+\text{N}^a{\cal H}_a \Big)\delta\text{t}_f\nonumber\\
{} & + \int\limits_{\text{B}_{f}}\text{dx}^2(16\pi)^{-1}\eta\ \delta \sqrt{\overline{\sigma}} \nonumber\\
{} & -\int\limits_{\text{B}_{f}}\text{dx}^2\Big(\overline{\text{N}}\ \overline{{\cal H}}+\overline{\text{N}}^j\ \overline{{\cal H}}_j \Big)\delta\text{t}_f\Bigg]\text{e}^{\text{i}\text{S}}.
\end{align}
Taking into account that $\delta\sqrt{\overline{\sigma}}_{|_{\text{B}_{f}}}=0$, and defining the proper time and proper space on $\text{M}$, and $\text{B}$, as
\bea\label{propers}
\delta\tau_f & = & \text{N}\delta\text{t}_f \quad \ ; \quad \delta\overline{\tau}_f = \overline{\text{N}}\delta\text{t}_f,\nonumber\\
\delta\chi^a & = & \text{N}^a\delta\text{t}_f \quad ; \quad \delta\overline{\chi}^j = \overline{\text{N}}^j\delta\text{t}_f,
\eea
from \eqref{psi-variation} we get that on $\Sigma_{f}$,
\bea\label{momenta-wf}
-\text{i}\frac{\delta}{\delta \Lambda}\int \text{D}\mu\text{e}^{\text{i}\text{S}} & = & \int \text{D}\mu\ \text{P} \ \text{e}^{\text{i}\text{S}},\nonumber\\
-\text{i}\frac{\delta}{\delta \Lambda^i}\int \text{D}\mu\text{e}^{\text{i}\text{S}} & = & \int \text{D}\mu\ \text{P}_i \ \text{e}^{\text{i}\text{S}},\nonumber\\
-\text{i}\frac{\delta}{\delta \sigma_{ij}}\int \text{D}\mu\text{e}^{\text{i}\text{S}} & = & \int \text{D}\mu\ \text{P}^{ij} \ \text{e}^{\text{i}\text{S}},
\eea
and
\bea\label{time-dep-M}
\frac{\partial\Psi}{\partial\tau_f} & = & -\text{i}\int\limits_{\Sigma_{f}}\text{dr}\text{d}^2x\int\text{D}\mu \ {\cal H} \ \text{e}^{\text{i}\text{S}}, \nonumber\\
\frac{\partial\Psi}{\partial\chi^a} & = & -\text{i}\int\limits_{\Sigma_{f}}\text{dr}\text{d}^2x\int\text{D}\mu \ {\cal H}_a \ \text{e}^{\text{i}\text{S}},
\eea
and
\bea\label{Sch-b-e}
\frac{\partial\Psi}{\partial\overline{\tau}_f} & = & -\text{i}\int\limits_{\text{B}_{f}}\text{d}^2x\int\text{D}\mu \ \overline{{\cal H}} \ \text{e}^{\text{i}\text{S}},\nonumber\\
\frac{\partial\Psi}{\partial\overline{\chi}^j} & = & -\text{i}\int\limits_{\text{B}_{f}}\text{d}^2x\int\text{D}\mu \ \overline{{\cal H}}_j \ \text{e}^{\text{i}\text{S}}.
\eea

\subsection{Constraints}

The functional integral defining the wave function \eqref{wf} contains integrals over $\text{N}$, and $\text{N}^a$,  see the discussion in the beginning of this section. The value of the integral should be
left unchanged by an infinitesimal translation of the integration variables $\text{N}$, and $\text{N}^a$, \cite{Hartle:1983ai}. Namely,
\be
\int \text{D}\mu \Big[\frac{\delta \text{S}}{\delta \text{N}}\Big]\text{e}^{\text{i}\text{S}}=\int \text{D}\mu  \ {\cal H}\big[\Lambda,\Lambda^i,\sigma_{ij},\text{P},\text{P}_i,\text{P}^{ij}\big] \ \text{e}^{\text{i}\text{S}}=0,
\ee
and
\be
\int \text{D}\mu \Big[\frac{\delta \text{S}}{\delta \text{N}^a}\Big]\text{e}^{\text{i}\text{S}}=\int \text{D}\mu \  {\cal H}_a\big[\Lambda,\Lambda^i,\sigma_{ij},\text{P},\text{P}_i,\text{P}^{ij}\big] \ \text{e}^{\text{i}\text{S}}=0.
\ee
Now using \eqref{momenta-wf}, and evaluating the previous expressions at $\text{t}=\text{t}_f$, we get the quantum counterpart of the Hamiltonian and the momentum constraint
\be\label{wde1}
\int \text{D}\mu  \ {\cal H}\big[\Lambda,\Lambda^i,\sigma_{ij},\text{P},\text{P}_i,\text{P}^{ij}\big] \ \text{e}^{\text{i}\text{S}}={\cal H}\big[\Lambda,\Lambda^i,\sigma_{ij},-\text{i}\frac{\delta}{\delta \Lambda},-\text{i}\frac{\delta}{\delta \Lambda^i},-\text{i}\frac{\delta}{\delta \sigma_{ij}}\big]\Psi=0,
\ee
and
\be\label{wde2}
\int \text{D}\mu \  {\cal H}_a\big[\Lambda,\Lambda^i,\sigma_{ij},\text{P},\text{P}_i,\text{P}^{ij}\big]\ \text{e}^{\text{i}\text{S}}={\cal H}_a\big[\Lambda,\Lambda^i,\sigma_{ij},-\text{i}\frac{\delta}{\delta \Lambda},-\text{i}\frac{\delta}{\delta \Lambda^i},-\text{i}\frac{\delta}{\delta \sigma_{ij}}\big]\Psi=0.
\ee
These are known in the QG literature as Wheeler–DeWitt equation and momentum constraint.
Notice that these two constraints imply that
\be
\frac{\partial\Psi}{\partial\tau_f} =\frac{\partial\Psi}{\partial\chi^a}=0,
\ee
which follows from \eqref{time-dep-M}.

We could try to proceed in a similar way for the boundary equations \eqref{Sch-b-e} however, the form of the functions $\overline{\cal H}$, and $\overline{\cal H}_j$, obstructs us to do that. Nonetheless, we can rewrite these two relations in a simpler form as
\bea\label{Sch-b-e1}
\frac{\partial\Psi}{\partial\overline{\tau}_f} & = & \int\limits_{\text{B}_{f}}\text{d}^2x \ \frac{\delta \Psi}{\delta\overline{\text{N}}},\nonumber\\
\frac{\partial\Psi}{\partial\overline{\chi}^j} & = & \int\limits_{\text{B}_{f}}\text{d}^2x \  \frac{\delta \Psi}{\delta\overline{\text{N}}^j}.
\eea

In addition to these two simpler equations there is a constraint over $\text{B}$. As in the previous discussion of the Hamiltonian and momentum constraint, the path integral defining the wave function \eqref{wf} contains also and integral over $\eta$, see for instance Table. \ref{table666},   and the value of the integral should be
left unchanged by an infinitesimal translation of this integration variable. More precisely,
\be
\int \text{D}\mu \Big[\frac{\delta \text{S}}{\delta \eta}\Big]\text{e}^{\text{i}\text{S}}=0,
\ee
or
\be\label{q-boundary_eom}
\int \text{D}\mu\Big[\text{P}\ \text{cosh}(\eta)-(8\pi)^{-1} \sqrt{\overline{\sigma}} {}^{(2)}\text{K}\ \text{sinh}(\eta)
 -(8\pi)^{-1}\sqrt{\overline{\sigma}} {}^{(2)}\overline{\text{K}}\Big]\text{e}^{\text{i}\text{S}}=0,
\ee
this constraint equation is just the quantum counterpart of the boundary equation of motion \eqref{boundary_eom_compact}. Specializing \eqref{q-boundary_eom} at $\text{t}=\text{t}_f$, it can be written as
\be\label{q-boundary_eom1}
\text{i} \frac{\delta}{\delta \Lambda}\Big\langle \text{cosh}(\eta)\Big\rangle+(8\pi)^{-1} \sqrt{\overline{\sigma}} {}^{(2)}\text{K}\ \Big\langle\text{sinh}(\eta)\Big\rangle
 +(8\pi)^{-1}\sqrt{\overline{\sigma}} {}^{(2)}\overline{\text{K}}\ \Psi=0,
\ee
where
\be
\Big\langle\text{...}\Big\rangle=\int\text{D}\mu(\text{...})\text{e}^{\text{i\text{S}}}.
\ee
We stress  that all the functions in \eqref{q-boundary_eom1} are evaluated on  $\text{B}_f=\text{J}_f$.

\section{Conclusions}\label{sec:5}

This work has pointed out a subtle fact about classical and quantum gravity on manifolds with boundaries. It turns out that on a manifold with spatial boundaries, some of the ADM degrees of freedom that are independent on the bulk, on the timelike boundary, they are not \eqref{parametrization1} and \eqref{parametrization2}.
This fact leads to a boundary equation of motion \eqref{boundary_eom_compact} derived from the Hamiltonian form of the action \eqref{Baction}.

The emergence of a classical equation of motion on the boundary in the Hamiltonian formulation of gravity seems to contradict the Lagrangian formulation \eqref{totalaction} where no boundary equation arises
\eqref{eom}. However, this is not the case, and no contradiction arises because this classical boundary equation turns out to be an identity when evaluated on a solution of the Einstein's equations \eqref{momenta-pi-ab}-\eqref{derivativeofdet}.

At the quantum level, the situation changes radically regarding the paradigm in QG on manifolds with boundaries. This new classical equation (identity) becomes a constraint equation on the boundary \eqref{q-boundary_eom1} similar to the Hamiltonian and the momentum constraints on the bulk. Thus, the time evolution of the wave function now is ruled by a Schr\"{o}dinger like equation depending only on the degrees of freedom on the junction $\text{B}_f=\text{J}_f$, and by this new constraint equation. This new equation certainly will have consequences when studying the time evolution in QG.

Along this work we have used (reproduced) well known results from the literature  related to QG on manifolds with boundaries. They have been essential for presenting this work in a rigorous and self-consistent manner. However, at this point it is worth to emphasize why we have been able to go further than previous works.

Previous works regarding this particular topic in QG on manifolds with boundaries focused mainly in getting either the boundary Schr\"{o}dinger equation, see for instance \cite{Hayward:1992ix, Feng:2017xsh},  or the boundary Hamiltonian \cite{Hawking:1996ww}; as we also have done here. Nonetheless,  they did not look for the classical equations of motion.

One of the main differences  with previous work is that they lacked of an exhaustive analysis of the boundary conditions over the spatial boundary. Like the one presented in section \ref{sec:3} yielding to equations \eqref{parametrization1} and \eqref{parametrization2}.

Another difference is that we have made the observation that the lapse and the shift functions are not completely fixed over the spatial boundary. A direct consequence of this observation reflects in the ADM Hamiltonian formalism. In this formalism now one encounters a very interesting and new mathematical problem. Now, to get the Hamiltonian and the momentum constraints, one has to vary $\text{N}$, and $\text{N}^{\text{r}}$ (which are Lagrange multipliers in the ADM formalism), in the bulk, but because of these fields are not fixed over $\text{B}$, one has to vary them over $\text{B}$ as well. Over $\text{B}$, however, these two Lagrange multipliers are not independent, but their dependence can be parameterized with the scalar field $\eta$. Up to the knowledge of the author this is the first time in the literature where this kind of mathematical problem is stated.

This parametrization was discussed in sections \ref{sec:3.1} and \ref{sec:3.2}, and is what later allowed us to find a classical boundary equation of motion and its quantum (constraint equation) counterpart.

On the one hand, these two differences with pervious works exposed above together with the classical boundary equation of motion \eqref{boundary_eom_compact}  and the (non-trivial) proof that it is an on shell identity are the new contributions of this work. On the other hand, the main and new result presented here is the constraint equation \eqref{q-boundary_eom} or  \eqref{q-boundary_eom1}. This constraint equation complements the boundary  Schr\"{o}dinger equation \eqref{Sch-b-e1} and, together with the Wheeler-Dewitt equation \eqref{wde1} and the momentum constraint \eqref{wde2} in the bulk, rule the time evolution of a QG state.

Operatively equation \eqref{q-boundary_eom1} does not look easy to solve. We do not know yet how to implement this constraint in general. Perhaps for some midisuperspace model, like the spherically symmetric one \cite{Kuchar:1994zk}, with a spatial boundary (the problem over the spatial boundary becomes an ordinary quantum mechanics), one can implement it. Or maybe in some two-dimensional models of gravity, like dilaton gravity. JT gravity \cite{Jackiw:1984je,Teitelboim:1983ux}, \cite{Benedict:1996qy}, and CGHS model \cite{Callan:1992rs}, see also \cite{Grumiller:2002nm} and references therein. In these cases, the problem over the spatial boundary would be a one-dimensional quantum mechanics problem. Another setup where we could implement this constraint could be in studying perturbations around a classical background in two \cite{Giddings:2021ipt}, or four dimensions \cite{Hayward:1992ix}, see also \cite{Esposito:1993ak} and references there in. Nonetheless, it is worth to remark that given the simple dependence of the boundary action on $\eta$, the $\eta$ integral in the wave function definition \eqref{wf} can be pointwise performed over the spatial boundary \cite{rosabal} giving rise to modified Bessel functions of the second kind or Hankel functions.

One of the main remaining questions is whether this new constraint equation will allow the Schr\"{o}dinger evolution of a QG state. For example, could it happen that this constraint restricts the time evolution of the wave function on the boundary to those solutions that do not depend on time, in a similar way the Hamiltonian or the momentum constraint do in the bulk. This question will be answered elsewhere.

\vspace{1cm}

\section*{Acknowledgments}

We are grateful to  Pablo Diaz for discussions and valuable comments. We would especially like to thank Jose Juan Fernandez-Melgarejo for inviting me to present this work at the University of Murcia and the Physics department members for the stimulating discussion during the presentation.  We also especially  thank Giampiero Esposito for his comments and for confirming the new results presented here.

\appendix

\section{\label{appen:A}Metrics matrix form}

\be
\text{g}_{\mu\nu}  = \begin{pmatrix}
                               -(\text{N}^2-\text{h}_{ab}\text{N}^a\text{N}^b) & \text{h}_{cb}\text{N}^c \\
                               \text{h}_{ac}\text{N}^c & \text{h}_{ab}
                             \end{pmatrix},
\ee

\be
\text{g}^{\mu\nu}  = \begin{pmatrix}
                               -\text{N}^{-2} & \text{N}^{-2}\text{N}^b \\
                               \text{N}^{-2}\text{N}^a & (\text{h}^{ab}-\text{N}^{-2}\text{N}^a\text{N}^b)
                             \end{pmatrix},
\ee

\be
\text{h}_{ab}  = \begin{pmatrix}
                               (\Lambda^2+\sigma_{ij}\Lambda^i\Lambda^j) & \sigma_{kj}\Lambda^k \\
                               \sigma_{ik}\Lambda^k & \sigma_{ij}
                             \end{pmatrix},
\ee

\be
\text{h}^{ab}  = \begin{pmatrix}
                               \Lambda^{-2} & -\Lambda^{-2}\Lambda^j \\
                               -\Lambda^{-2}\Lambda^i & (\sigma^{ij}+\Lambda^{-2}\Lambda^i\Lambda^j)
                             \end{pmatrix},
\ee

\be
\text{g}^{\mu\nu}  = \begin{pmatrix}
                               -\text{N}^{-2} & \text{N}^{-2}\text{N}^r & \text{N}^{-2}\text{N}^j \\
                               \text{N}^{-2}\text{N}^r & (\Lambda^{-2}-\text{N}^{-2}\text{N}^r\text{N}^r) & (-\Lambda^{-2}\Lambda^j-\text{N}^{-2}\text{N}^r\text{N}^j)\\
                               \text{N}^{-2}\text{N}^i & (-\Lambda^{-2}\Lambda^i-\text{N}^{-2}\text{N}^r\text{N}^i) & (\sigma^{ij}+\Lambda^{-2}\Lambda^j\Lambda^j-\text{N}^{-2}\text{N}^i\text{N}^j)
                             \end{pmatrix}.
\ee

\section{\label{appen:B}Extrinsic Curvature}

The extrinsic curvature of a surface embedded in a higher dimensional space is given by
\be
\text{K}=-\nabla_{\mu} \hat{\text{n}}^{\mu},
\ee
where $\hat{\text{n}}^{\mu}$, is the unit normal vector to the surface. For the case in which the surface is the two-dimensional boundary $\text{B}_{\text{t}}$, embedded in $\Sigma_{\text{t}}$, the extrinsic curvature can be written as
\be
{}^{(2)}\text{K}=-\text{D}_{a} {}^{(3)}\hat{\text{r}}^{a},
\ee
where $\text{D}_{a}$, is the covariant derivative compatible with the metric $\text{h}_{ab}$, on $\Sigma_{\text{t}}$, and $\text{r}_a$, the unit normal vector (with respect to the three metric $\text{h}_{ab}$) to $\text{B}$.

Using the normal and tangential decomposition of the divergence of a vector
\be\label{n-t-deco}
\nabla_{\mu}\text{V}^{\mu}=\text{D}_a\text{v}^a+\hat{\text{n}}^{\mu}\text{V}^{\nu}\nabla_{\mu}\hat{\text{n}}_{\nu},
\ee
with
\be
\text{V}^{\mu}=\text{v}^a\gamma_a^{\mu},
\ee
where $\gamma_a^{\mu}$, is a basis of tangent vectors on $\Sigma_{\text{t}}$. In the coordinates $(\text{t},\text{r},\text{x}^1,\text{x}^2)$, these vectors (over a constant time surface $\text{t}=\text{t}_{0}$) take the simple form
\be\label{basis}
\gamma^{\mu}_a =  \delta^{\mu}_a\quad,\quad a=1,2,3.
\ee
Using \eqref{n-t-deco} it is straightforward to write the extrinsic curvature as
\be\label{b5}
{}^{(2)}\text{K}=-\text{g}^{\mu\nu}\nabla_{\mu}{}^{(3)}\hat{\text{r}}_{\nu}+
\hat{\text{n}}^{\mu}\nabla_{\mu}\hat{\text{n}}_{\nu}{}^{(3)}\hat{\text{r}}^{\nu}.
\ee
where
\be
{}^{(3)}\hat{\text{r}}^{\mu}={}^{(3)}\hat{\text{r}}^a\gamma_a^{\mu}\label{emnormalaB},
\ee
is the unit normal vector to $\text{B}_{\text{t}}$, (with respect to the metric $\text{h}_{ab}$) embedded in the four-dimensional space.

\section{\label{appen:C}Algebraic Manipulation to get the Boundary Action}

Our starting point is the gravity action  with the boundary contributions as well as the junction contribution \eqref{totalaction}.
Using \eqref{R3} and $\sqrt{-g}=\text{N}\sqrt{\text{h}}$, and integrating the total derivatives we get
\begin{align}\label{c3}
  \text{S} & = \frac{1}{16 \pi} \int \text{dt} \int\limits_{\Sigma_{\text{t}}}\text{d}^3x \text{N}\sqrt{\text{h}}\Big( \text{K}_{ab}\text{K}^{ab}-\text{K}^2+ {}^{(3)}\text{R}\Big)\nonumber \\
  {} & +\frac{1}{8 \pi} \int\limits_{\Sigma_f-\Sigma_i}\text{d}^3y \sqrt{\text{h}}\ \text{K}-\frac{1}{8 \pi} \int\limits_{\text{B}}\text{d}^3\overline{y} \sqrt{-\overline{\text{h}}}\ \overline{\text{K}}+\frac{1}{8 \pi} \int\limits_{\text{J}_f-\text{J}_i}\text{d}^2z \sqrt{\overline{\sigma}}\ \eta \nonumber\\
  {} & -\frac{1}{8 \pi} \int\limits_{\Sigma_f-\Sigma_i}\text{d}^3y \sqrt{\text{h}} \ \hat{\text{n}}_{\mu}\big(-\text{K}\hat{\text{n}}^{\mu}-\hat{\text{n}}^{\nu}\nabla_{\nu}\hat{\text{n}}^{\mu} \big)\nonumber\\
  {} & +\frac{1}{8 \pi} \int\limits_{\text{B}}\text{d}^3y \sqrt{-\overline{\text{h}}} \ \hat{\text{r}}_{\mu}\big(-\text{K}\hat{\text{n}}^{\mu}-\hat{\text{n}}^{\nu}\nabla_{\nu}\hat{\text{n}}^{\mu} \big),
\end{align}
where $\hat{\text{r}}_{\mu}$, is the four-dimensional unit normal vector \eqref{refcapri1},
with $\text{M}=\big(\text{N}^2-(\Lambda\text{N}^{\text{r}})^2\big)^{\frac{1}{2}}$, and $\hat{\text{n}}_{\mu}$, is the four-dimensional unit normal vector to $\Sigma_t$, denoted in section \ref{sec:3.1} as $\hat{\text{n}}_{\mu}^{(\Sigma)}$. Recall that on \text{B}, $\text{M}=\big(\text{N}^2-(\Lambda\text{N}^{\text{r}})^2\big)^{\frac{1}{2}}_{|_{\text{B}}}=\overline{\text{N}}$. For simplicity we have omitted the background contribution, at the end of the calculation it can be easily  reinserted.

Now taking into account that $\hat{\text{n}}_{\mu}\hat{\text{n}}^{\mu}=-1$, and $\nabla_{\nu}\hat{\text{n}}^{\mu}\ \hat{\text{n}}_{\mu}=0$, we see that the integrals $\int\limits_{\Sigma_f-\Sigma_i}$ cancel out to get
\begin{align}\label{c5}
  \text{S} & = \text{S}_{\text{M}}+\text{S}_{\text{B}}\nonumber\\
  {} & =\frac{1}{16 \pi} \int \text{dt} \int\limits_{\Sigma_{\text{t}}}\text{d}^3x \text{N}\sqrt{\text{h}}\Big( \text{K}_{ab}\text{K}^{ab}-\text{K}^2+ {}^{(3)}\text{R}\Big)\nonumber \\
  {} & -\frac{1}{8 \pi} \int\limits_{\text{B}}\text{d}^3\overline{y} \sqrt{-\overline{\text{h}}}\big(-\text{g}^{\mu\nu}\nabla_{\mu}\hat{\text{r}}_{\nu}
  +\hat{\text{r}}_{\mu}\hat{\text{n}}^{\nu}\nabla_{\nu}\hat{\text{n}}^{\mu} \big)\nonumber\\
  {} & -\frac{1}{8 \pi} \int\limits_{\text{B}}\text{d}^3\overline{y} \sqrt{-\overline{\text{h}}}\ \hat{\text{r}}_{\mu} \hat{\text{n}}^{\mu} \text{K}+\frac{1}{8 \pi} \int\limits_{\text{J}_f-\text{J}_i}\text{d}^2z \sqrt{\overline{\sigma}}\ \eta.
\end{align}

Using the decomposition \eqref{refcapri2}
and
\be
\hat{\text{r}}_{\mu}\hat{\text{n}}^{\mu}=\text{M}^{-1}\Lambda \text{N}^{\text{r}},
\ee
where ${}^{(3)}\hat{\text{r}}_{\mu}=\Lambda(-\text{N}^{\text{r}},-1,0,0)$. Notice that $\hat{\text{n}}^{\mu}=\text{N}^{-1}(1,-\text{N}^{\text{r}},-\text{N}^j)$, which implies ${}^{(3)}\hat{\text{r}}_{\mu}\hat{\text{n}}^{\mu}=0$,
with ${}^{(3)}\hat{\text{r}}^{\mu}$,  \eqref{emnormalaB} being the unit normal vector to $\text{B}_{\text{t}}$ (with respect to the metric $\text{h}_{ab}$, and at the same time it is tangent to $\Sigma_{\text{t}}$) seen from the four-dimensional space,  and  $\hat{\text{r}}^a=\text{h}^{ab}\hat{\text{r}}_b$, with
\be\label{c8}
{}^{(3)}\hat{\text{r}}_a=\Lambda(-1,0,0).
\ee
We arrive at
\begin{align}\label{c9}
\text{S}_{\text{B}} & =  -\frac{1}{8 \pi} \int\limits_{\text{B}}\text{d}^3\overline{y} \sqrt{-\overline{\text{h}}} \text{N}\text{M}^{-1}\big(-\text{g}^{\mu\nu}\nabla_{\mu}{}^{(3)}\hat{\text{r}}_{\nu}
  +{}^{(3)}\hat{\text{r}}_{\mu}\hat{\text{n}}^{\nu}\nabla_{\nu}\hat{\text{n}}^{\mu} \big)\nonumber\\
  {} & -\frac{1}{8 \pi} \int\limits_{\text{B}}\text{d}^3\overline{y} \sqrt{-\overline{\text{h}}}\Big(-{}^{(3)}\hat{\text{r}}^{\mu}\partial_{\mu}(\text{N}\text{M}^{-1})
+\hat{\text{n}}^{\mu}\partial_{\mu}(\Lambda\text{N}^{\text{r}}\text{M}^{-1}) \Big)
 +\frac{1}{8 \pi} \int\limits_{\text{J}_f-\text{J}_i}\text{d}^2z \sqrt{\overline{\sigma}}\ \eta.
\end{align}
Now using $\sqrt{-\overline{\text{h}}}=\overline{\text{N}}\sqrt{\overline{\sigma}}$, and $\text{M}_{|_{\text{B}}}=\overline{\text{N}}$, and \eqref{b5} we get
\begin{align}\label{c10}
\text{S}_{\text{B}} & =  -\frac{1}{8 \pi} \int\limits_{\text{B}}\text{d}^3\overline{y}\ \text{N}\ \sqrt{\overline{\sigma}}\ {}^{(2)}\text{K}\nonumber\\
  {} & -\frac{1}{8 \pi} \int\limits_{\text{B}}\text{d}^3\overline{y}\  \overline{\text{N}}\ \sqrt{\overline{\sigma}}\Big(-{}^{(3)}\hat{\text{r}}^{\mu}\partial_{\mu}(\text{N}\text{M}^{-1})
+\hat{\text{n}}^{\mu}\partial_{\mu}(\Lambda\text{N}^{\text{r}}\text{M}^{-1}) \Big)\nonumber\\
{} &  +\frac{1}{8 \pi} \int\limits_{\text{J}_f-\text{J}_i}\text{d}^2z \sqrt{\overline{\sigma}}\ \eta.
\end{align}

Using the boundary conditions \footnote{Some times during the subsequent calculation it is better to reexpress the boundary conditions as
\be
\text{N}=\overline{\text{N}}\Big(1+(\frac{\Lambda \text{N}^{\text{r}}}{\overline{\text{N}}})^2 \Big)^{\frac{1}{2}}\nonumber,
\ee
and
\be
\text{N}^i=\overline{\text{N}}^i-\Lambda^i\text{N}^{\text{r}}\nonumber.
\ee
}
\eqref{parametrization1} and \eqref{parametrization2} and
\be
{}^{(3)}\hat{\text{r}}^{\mu}=(0,-\Lambda^{-1},\Lambda^{-1}\Lambda^j),
\ee
together with \eqref{eta_def1}, we can show that the second and third line of $\eqref{c10}$ can be written as
\be\label{c11}
\frac{1}{8 \pi} \int\limits_{\text{B}}\text{d}^3\overline{y} \big(\partial_{\text{t}}(\sqrt{\overline{\sigma}})-\partial_j(\sqrt{\overline{\sigma}} \ \overline{\text{N}}^j) \big)\ \eta.
\ee
To get \eqref{c11} we have integrated by parts and considered that $\text{B}_t$ is a compact manifold, which of course, it is always the case. Notice that \eqref{c11} is defined over the whole timelike boundary and not only over the junctions. Also the scalar field $\eta$, in \eqref{c11} coincides with  \eqref{eta}, defined over $\text{B}$, and not only over the junctions. In fact after the integration by parts the junctions contributions in \eqref{c10} cancel out.

Summarizing, the Lagrangian form of the boundary action, using \eqref{eta_def1} and \eqref{extrincBtinB}, can be written as
\be\label{c12}
  \text{S}_{\text{B}}= - \frac{1}{8 \pi} \int\text{dt}\int\limits_{\text{B}_{\text{t}}}\text{dx}^2\ \overline{\text{N}} \sqrt{\overline{\sigma}}\Big( {}^{(2)}\text{K}\ \text{cosh}(\eta)+\overline{\text{K}}\ \eta\Big).
\ee


\newpage

\begin{thebibliography}{99}


\bibitem{Almheiri:2019hni}
A.~Almheiri, R.~Mahajan, J.~Maldacena and Y.~Zhao,
``The Page curve of Hawking radiation from semiclassical geometry,''
JHEP \textbf{03}, 149 (2020),
[arXiv:1908.10996 [hep-th]].


\bibitem{Penington:2019kki}
G.~Penington, S.~H.~Shenker, D.~Stanford and Z.~Yang,
``Replica wormholes and the black hole interior,''
[arXiv:1911.11977 [hep-th]].


\bibitem{Almheiri:2019qdq}
A.~Almheiri, T.~Hartman, J.~Maldacena, E.~Shaghoulian and A.~Tajdini,
``Replica Wormholes and the Entropy of Hawking Radiation,''
JHEP \textbf{05}, 013 (2020), [arXiv:1911.12333 [hep-th]].


\bibitem{Giddings:2021ipt}
S.~B.~Giddings,
``Schr\"odinger evolution of two-dimensional black holes,''
[arXiv:2108.07824 [hep-th]].


\bibitem{Hawking:1976ra}
S.~W.~Hawking,
``Breakdown of Predictability in Gravitational Collapse,''
Phys. Rev. D \textbf{14}, 2460-2473 (1976).

\bibitem{Hayward:1992ix}
G.~Hayward and K.~Wong,
``Boundary Schrodinger equation in quantum geometrodynamics,''
Phys. Rev. D \textbf{46}, 620-626 (1992), Addendum Phys. Rev. D \textbf{47}, (1993).


\bibitem{Hayward:1993my}
G.~Hayward,
``Gravitational action for space-times with nonsmooth boundaries,''
Phys. Rev. D \textbf{47}, 3275-3280 (1993).


\bibitem{Feng:2017xsh}
J.~C.~Feng and R.~A.~Matzner,
``From path integrals to the Wheeler-DeWitt equation: Time evolution in spacetimes with a spatial boundary,''
Phys. Rev. D \textbf{96}, no.10, 106005 (2017),
[arXiv:1708.07001 [gr-qc]].

\bibitem{Hawking:1996ww}
S.~W.~Hawking and C.~J.~Hunter,
``The Gravitational Hamiltonian in the presence of nonorthogonal boundaries,''
Class. Quant. Grav. \textbf{13}, 2735-2752 (1996),
[arXiv:gr-qc/9603050 [gr-qc]].



\bibitem{Brown:2000dz}
J.~D.~Brown, S.~R.~Lau and J.~W.~York, Jr.,
``Action and energy of the gravitational field,''
[arXiv:gr-qc/0010024 [gr-qc]].


\bibitem{DeWitt:1967yk}
B.~S.~DeWitt,
``Quantum Theory of Gravity. 1. The Canonical Theory,''
Phys. Rev. \textbf{160}, 1113-1148 (1967).


\bibitem{Arnowitt:1959eec}
R.~Arnowitt and S.~Deser,
``Quantum Theory of Gravitation: General Formulation and Linearized Theory,''
Phys. Rev. \textbf{113}, 745-750 (1959).


\bibitem{Arnowitt:1959ah}
R.~L.~Arnowitt, S.~Deser and C.~W.~Misner,
``Dynamical Structure and Definition of Energy in General Relativity,''
Phys. Rev. \textbf{116}, 1322-1330 (1959).


\bibitem{Arnowitt:1960es}
R.~L.~Arnowitt, S.~Deser and C.~W.~Misner,
``Canonical variables for general relativity,''
Phys. Rev. \textbf{117}, 1595-1602 (1960).


\bibitem{Gourgoulhon:2007ue}
E.~Gourgoulhon,
``3+1 formalism and bases of numerical relativity,''
[arXiv:gr-qc/0703035 [gr-qc]].

\bibitem{Gibbons:1976ue}
G.~W.~Gibbons and S.~W.~Hawking,
``Action Integrals and Partition Functions in Quantum Gravity,''
Phys. Rev. D \textbf{15}, 2752-2756 (1977).


\bibitem{York:1986lje}
J.~York,
``Boundary terms in the action principles of general relativity,''
Found. Phys. \textbf{16}, 249-257 (1986).



\bibitem{Saharian:2003dr}
A.~A.~Saharian,
``On the energy momentum tensor for a scalar field on manifolds with boundaries,''
Phys. Rev. D \textbf{69}, 085005 (2004).


\bibitem{Hartle:1983ai}
J.~B.~Hartle and S.~W.~Hawking,
``Wave Function of the Universe,''
Phys. Rev. D \textbf{28}, 2960-2975 (1983).


\bibitem{Teitelboim:1981ua}
C.~Teitelboim,
``Quantum Mechanics of the Gravitational Field,''
Phys. Rev. D \textbf{25}, 3159 (1982).


\bibitem{Hawking:1985bk}
S.~W.~Hawking and D.~N.~Page,
``Operator Ordering and the Flatness of the Universe,''
Nucl. Phys. B \textbf{264}, 185-196 (1986)




\bibitem{Kuchar:1994zk}
K.~V.~Kuchar,
``Geometrodynamics of Schwarzschild black holes,''
Phys. Rev. D \textbf{50}, 3961-3981 (1994),
[arXiv:gr-qc/9403003 [gr-qc]].


\bibitem{Callan:1992rs}
C.~G.~Callan, Jr., S.~B.~Giddings, J.~A.~Harvey and A.~Strominger,
``Evanescent black holes,''
Phys. Rev. D \textbf{45}, no.4, R1005 (1992),
[arXiv:hep-th/9111056 [hep-th]].


\bibitem{Jackiw:1984je}
R.~Jackiw,
``Lower Dimensional Gravity,''
Nucl. Phys. B \textbf{252}, 343-356 (1985).


\bibitem{Teitelboim:1983ux}
C.~Teitelboim,
``Gravitation and Hamiltonian Structure in Two Spacetime Dimensions,''
Phys. Lett. B \textbf{126}, 41-45 (1983)


\bibitem{Benedict:1996qy}
E.~Benedict, R.~Jackiw and H.~J.~Lee,
``Functional Schrodinger and BRST quantization of (1+1)-dimensional gravity,''
Phys. Rev. D \textbf{54}, 6213-6225 (1996),
[arXiv:hep-th/9607062 [hep-th]].


\bibitem{Grumiller:2002nm}
D.~Grumiller, W.~Kummer and D.~V.~Vassilevich,
``Dilaton gravity in two-dimensions,''
Phys. Rept. \textbf{369}, 327-430 (2002),
[arXiv:hep-th/0204253 [hep-th]].



\bibitem{Esposito:1993ak}
G.~Esposito,
``Canonical and perturbative quantum gravity,''
SISSA-10-93-A.

\bibitem{rosabal}

J.~A.~Rosabal, ``Schr\"{o}dinger Evolution and Constraints in JT Gravity on the Strip,'' work in progress.

\end{thebibliography}
\end{document}